# Conditional statistical physical properties in two-joint complex systems having long-range interactions

Zhifu Huang*

College of Information Science and Engineering, Huaqiao University, Xiamen 361021, People's Republic of China

We propose the sum and the difference of the normalized velocity of two-joint systems to describe its long-range interaction. It is found that the conditional probability distribution function (CPDF) of the normalized velocity between two-joint systems can be derived. The analytical CPDF needs only six parameters for arbitrary two-joint systems. Two typical currency exchange databases, i.e., EUR/USD and GBP/USD, which collect the minutely opening exchange prices from 1 January 1999 to 31 December 2011, are adopted as example. By calculating the CPDF in the currency exchange databases, it is shown that all of the results are well described by our theory. We also use the analytical CPDF to calculate the entropy of two-joint systems, it is found that the entropy of two-joint systems is less than the sum of entropy of each system in the two currency exchange databases. It means that some information of two-joint systems may overlap. We must important to note that the results presented here do not need to know the form of interaction of two-joint systems, and the analytical CPDF is a general expression. It is valid not only for current exchange systems, but also helpful for investigate other conditional statistical properties in any two-joint long-range interactions complex systems.

**Key words**: Complex systems; Long-range interaction; Conditional probability distribution function

**PACS numbers**: 05.65.+b; 05.45.Ac; 02.50.Cw

________________________

28    *Email: zfhuang@hqu.edu.cn

The relationship of the whole and its part is an important theme in science. The core of this problem is to obtain the condition statistical physical properties of the whole and its parts. However, because of the complexity of their interactions, the condition statistical physical properties of joint systems are largely unknown. In the present work, we will introduce a new method to obtain the condition statistical physical properties of two-joint systems which do not need to know the interactions form of two-joint systems. First of all, in order to obtain the conditional statistical physical properties of the whole and its parts, we must analyze the probability distribution function (PDF) from real complex systems. The central limit theorem (CLT) is an extremely important concept in probability theory and also lies at the heart of statistical physics [1], [2]. It basically says that the sum of $N$ independent identically distributed random variables, rescaled with a factor $1/\sqrt{N}$, has a Gaussian distribution. However, the PDF of velocity in long-range interaction complex systems usually do not agree with Gaussian distribution coming from independent or nearly independent contributions, but may take the forms of non-Gaussian distribution, e.g., Levy stable forms [3]-[6], stretched Gaussian shape [6], [7], or the form of q-Gaussian given by

$$p(y) = C[1-(1-q)\beta y^2]^{1/(1-q)}, \qquad (1)$$

where $\beta$ is a parameter characterizing the width of the distribution and q is the nonextensivity index [8], [9], while C is the normalized parameter. In Eq. (1), $q \neq 1$ indicates a departure from the Gaussian shape while $q \to 1$ limit yields the normal Gaussian distribution. When $q > 1$ and $y^2 >> 1/[(q-1)\beta]$, equation (1) may be simplified into the power-law shape $p(y) \propto y^\gamma$, where $\gamma = 2/(1-q)$. It means that the q-Gaussian form of PDF may be used to analyze the asymptotic of Lévy stable forms [10]. On the other hand, the normalized parameter C depend on $\beta$ and q as

$$C = \frac{\sqrt{\beta(q-1)}\Gamma[1/(q-1)]}{\sqrt{\pi}\Gamma[(3-q)/(2q-2)]}, \quad (2)$$

where $\Gamma(...)$ is the gamma function, while q must in the range of $1 < q < 3$ to validate the gamma function. In addition, the variance of equation (1) also depend on $\beta$ and q as

$$\sigma^2 = \frac{\Gamma[(5-3q)/(2q-2)]}{2\beta(q-1)\Gamma[(3-q)/(2q-2)]}, \quad (3)$$

where q must in the range of $1 < q < 1.6$ to keep the gamma function valid. From equations (2) and (3), it can be seen that both C and $\sigma$ are determinate if $\beta$ and q are given. Mathematically speaking, there are only two independent parameters among $\beta$, q, C, and $\sigma$, so that we can choose arbitrary two among them to describe the PDF. For example, we can choose q and $\sigma$ to describe the PDF of velocity in systems.

In real systems, the time interval of any processes is finite. When the position is ensured, the velocity during the time interval $\Delta t$ can be expressed as $v(t, \Delta t) = [R(t + \Delta t) - R(t)]/\Delta t$, where $R(t)$ is the position at time $t$ and $\Delta t$ is the time interval. On the other hand, in order to describe two-joint complex systems, the velocity in different systems can be normalized to the standard deviation as $u(t, \Delta t) = v(t, \Delta t)/\sigma_{v(t,\Delta t)}$. After that, when analyzing the relationship of two-joint complex systems, we can define two variables as

$$\psi(t, \Delta t) = \frac{u_1(t, \Delta t) + u_2(t, \Delta t)}{2}, \quad (4)$$

and

$$\theta(t,\Delta t) = \frac{u_2(t,\Delta t) - u_1(t,\Delta t)}{2}. \qquad (5)$$

Although the interaction form of two-joint systems may be very complex, we can find that the variables $\psi(t,\Delta t)$ and $\theta(t,\Delta t)$ include the normalized velocity from two-joint complex systems and may be used to analyze the interaction of two-joint systems. As a result, we may obtain conditional statistical properties of two-joint complex systems from analyzing the relationship of variables $\psi(t,\Delta t)$ and $\theta(t,\Delta t)$. Using equations (4) and (5), one can find that when the event $u_1(t,\Delta t)=y$ and $u_2(t,\Delta t)=z$ is ensured, the event $\psi(t,\Delta t)=(y+z)/2$ and $\theta(t,\Delta t)=(z-y)/2$ is also ensured at the same time. As a result, the joint probability distribution function (JPDF) of event $u_1(t,\Delta t)=y$ and $u_2(t,\Delta t)=z$ is equal to the JPDF of event $\psi(t,\Delta t)=(y+z)/2$ and $\theta(t,\Delta t)=(z-y)/2$. Then we can obtain

$$p_{u_1,u_2}(y,z) = p_{\psi,\theta}(\frac{y+z}{2}, \frac{z-y}{2}), \qquad (6)$$

where $p_{u_1,u_2}(y,z)$ is the JPDF of event $u_1(t,\Delta t)=y$ and $u_2(t,\Delta t)=z$, and $p_{\psi,\theta}[(y+z)/2,(z-y)/2]$ is the JPDF of event $\psi(t,\Delta t)=(y+z)/2$ and $\theta(t,\Delta t)=(z-y)/2$. It is well known that the JPDF of two events can be calculated as the PDF of the first event multiply the conditional probability distribution function (CPDF) of the next event in the condition of the first event happen. Then one can obtain the expression as

$$p_{u_1,u_2}(y,z) = p_{u_1}(y) p_{u_2|u_1}(z|y), \qquad (7)$$

and

$$p_{\psi,\theta}(y,z) = p_{\psi}(y) p_{\theta|\psi}(z|y), \qquad (8)$$

87  where $p_{u_1}(y)$ and $p_{\psi}(y)$ are the PDFs of event $u_1(t,\Delta t)=y$ and event

88  $\psi(t,\Delta t)=y$, respectively, while $p_{u_2|u_1}(z|y)$ is the CPDF of event $u_2(t,\Delta t)=z$ in the

89  condition of event $u_1(t,\Delta t)=y$, and $p_{\theta|\psi}(z|y)$ is the CPDF of event $\theta(t,\Delta t)=z$ in

90  the condition of event $\psi(t,\Delta t)=y$.

91  According to the Bayesian method, the prior CPDF or posterior CPDF can be

92  used to calculate the other CPDF. Substituting equations (7) and (8) into equation (6),

93  it is very natural to obtain the following relation

$$p_{u_2|u_1}(z|y)=p_{\psi}(\frac{y+z}{2})p_{\theta|\psi}(\frac{z-y}{2}|\frac{y+z}{2})/p_{u_1}(y). \tag{9}$$

94  From equation (9), we can calculate CPDF $p_{u_2|u_1}(z|y)$ from CPDF $p_{\theta|\psi}(z|y)$. On

95  the other hand, one can calculate the covariance of $\psi(t,\Delta t)$ and $\theta(t,\Delta t)$ from

96  equations (4) and (5) as

$$\langle\psi(t,\Delta t)\theta(t,\Delta t)\rangle=\frac{1}{4}[\langle u_2^2(t,\Delta t)\rangle-\langle u_1^2(t,\Delta t)\rangle]. \tag{10}$$

97  Fortunately, because we have normalized the velocity of two complex systems to the

98  standard deviation, equation (10) is equal to zero. Thus, we can obtain that the

99  incidence relation of $\psi(t,\Delta t)$ and $\theta(t,\Delta t)$ is neither positive nor negative.

100 Therefore, we can assume that the CPDF $p_{\theta|\psi}(z|y)$ is a symmetrical function. After

101 that, we may only analyze the conditional variance and tail of CPDF $p_{\theta|\psi}(z|y)$ to

102 obtain its expression. Since two-joint complex systems having long-range interaction,

103 the covariance of $u_1(t,\Delta t)$ and $u_2(t,\Delta t)$ may be not equal to zero. As a result, it is

104 possible and beneficial to obtain asymmetric CPDF $p_{u_2|u_1}(z|y)$ from symmetrical

105 CPDF $p_{\theta|\psi}(z|y)$ in two-joint long-range interaction complex systems.

106 The conditional statistical properties of currency market fluctuations are important for modeling and understanding complex market dynamics. We analyze two typical currency exchange databases of Euro vs U.S. Dollar (EUR/USD) and Great Britain Sterling Pound vs U.S. Dollar (GBP/USD), which can be found at the following websites: http://finance.yahoo.com and http://www.metaquotes.net. We collect the minutely opening exchange prices from 1 January 1999 to 31 December 2011, so the time interval $\Delta t$ is 1 minute for the data sequences. For the sake of convenience, we can normalize $\Delta t$ as 1 in our work. We also define one basic quantity [11]: the position $R(t)$ is the logarithmic of the exchange price $R(t) = \log[price(t)]$. After that, the corresponding normalized velocity can be obtained. Figure 1a show that the q-Gaussian distribution can be well approximated by the PDF of variables for the data, while different cases have different values of q.

We study two time series at the same moment, both of which can be considered as two outputs of a complex system. After that, we assume both time series can be seen as two-joint system. If there does not exist long-range interaction in the sequence of two-joint systems, the CPDF $p_{\theta|\psi}(z|y)$ is independent of $\psi(t,\Delta t)=y$. Otherwise, the dependence of CPDF $p_{\theta|\psi}(z|y)$ on $\psi(t,\Delta t)=y$ will reveal the long-range interaction within two-joint systems. Since the covariance of $\psi(t,\Delta t)$ and $\theta(t,\Delta t)$ is equal to zero, we can assume that the corresponding CPDF $p_{\theta|\psi}(z|y)$ also satisfies the *q*-Gaussian shape and it can be written as

$$p_{\theta|\psi}(z|y) = C_{\theta|\psi}(y)\{1-[1-q_{\theta|\psi}(y)]\beta_{\theta|\psi}(y)z^2\}^{1/[1-q_{\theta|\psi}(y)]}, \qquad (11)$$

where $C_{\theta|\psi}(y)$, $q_{\theta|\psi}(y)$ and $\beta_{\theta|\psi}(y)$ are conditional parameters which depend on $\psi(t,\Delta t)=y$. It should be mentioned that there are only two independent parameters in the system, so we can choose $\sigma^2_{\theta|\psi}(y)$ and $q_{\theta|\psi}(y)$ to analyze the CPDF $p_{\theta|\psi}(z|y)$, 129  where $\sigma^2_{\theta|\psi}(y)$ is the conditional variance that depends on $\psi(t,\Delta t)=y$ and $q_{\theta|\psi}(y)$

130  describes the fat-tail of the distribution. It can be seen from figure 1b that the

131  *q*-Gaussian distribution can be well approximated by a symmetrical CPDF $p_{\theta|\psi}(z|y)$.

132  For different values of y, the corresponding *q* values are also different. Thus, it is

133  possible to describe the CPDF $p_{\theta|\psi}(z|y)$ between two-joint long-range interaction

134  complex systems by *q*-Gaussian distributions.

135  Because complex systems sometimes exhibit long-range interaction, the

136  relationship of CPDF between the sum and the difference of normalized velocity may not

137  be easily obtained. However, an intriguing aspect of the complex systems is to

138  exhibit self-similar structures [12] characterized by scale invariance and plays a

139  central role in a large number of physics phenomena [13]. As the sum of normalized

140  velocity $\psi(t,\Delta t)$ include the difference of normalized velocity $\theta(t,\Delta t)$, we may

141  suppose that the sum of normalized velocity $\psi(t,\Delta t)$ the difference of normalized

142  velocity $\theta(t,\Delta t)$ in the same quasi-system may include self-similar structures. As a

143  result, the conditional variance between the sum of normalized velocity and the

144  difference of normalized velocity can be written as a linear shape

$$\sigma^2_{\theta|\psi}(y) = r_{\sigma 0} + r_{\sigma 1} y^2, \quad (12)$$

145  where $r_{\sigma 0}$ and $r_{\sigma 1}$ are the parameters of function $\sigma^2_{\theta|\psi}(y)$. Therefore, we can fit the

146  conditional variance with equation (12). As can be shown in figures 2a, equation (12)

147  can be well fitted by the data.

148  In addition, we can analyze the conditional q of the difference of normalized

149  velocity versus the sum of normalized velocity. As the covariance of $\psi(t,\Delta t)$ and

150  $\theta(t,\Delta t)$ equals zero, the conditional q may depend on the square of the sum of

151  normalized velocity. Furthermore, it should be mentioned that when the sum of normalized velocity is increased, the corresponding probability of the difference of normalized velocity will decrease and the corresponding conditional difference of normalized velocity will tend to being independent from each other. Thus, when the sum of normalized velocity is large enough, the conditional $q$ will tend to 1. For the sake of convenience, if the conditional $q$ is less than 1.01, we set it as 1.01. Therefore, the range of the sum of normalized velocity can not be too large and we fit the conditional $q$ values by the following expression,

$$q_{\theta|\psi}(y) = r_{q0} + r_{q1} y^2, \tag{13}$$

where $r_{q0}$ and $r_{q1}$ are two parameters of the function $p_{\theta|\psi}(z|y)$. It is shown in figure 2b that equation (13) is well fitted by the real data. Thus, we can find that the $q$-Gaussian distribution, conditional variance and conditional $q$ values are enough to describe the CPDF $p_{\theta|\psi}(z|y)$. Moreover, substituting equations (2), (3), (12) and (13) into equation (11), the CPDF $p_{\theta|\psi}(z|y)$ can be explicitly obtained, and equation (9) can be used to obtain CPDF $p_{u_2|u_1}(z|y)$. It is important to note that the CPDF $p_{u_2|u_1}(z|y)$ in all different cases can be explicitly expressed with only six parameters $q_\psi$, $\sigma_\psi$, $r_{\sigma 0}$, $r_{\sigma 1}$, $r_{q0}$ and $r_{q1}$, which can be obtained from the data fitting. It is shown clearly in figure 3 that the curves of theory are well approximated by the data in different cases, and CPDF $p_{u_2|u_1}(z|y)$ is asymmetric function when $y \neq 0$. It is implied that the JPDF of two-joint systems having long-range interaction and do not suite the independent principle of multiplying. It is also important to note that the analytical CPDF is a general expression, which may allow us to obtain the CPDF in other two-joint long-range interaction complex systems with the same method.

Furthermore, we may numerical calculate the conditional average of normalized velocity between two-joint systems by the obtained CPDF $p_{u_2|u_1}(z|y)$ as

$$M_{u_2|u_1}(y) = \int_{W_z|y} z p_{u_2|u_1}(z|y) dz, \qquad (14)$$

175  where $M_{u_2|u_1}(y)$ is the conditional average which depends on $u_1(t, \Delta t) = y$, while

176  $W_z|y$ means all situations in the condition $u_1(t, \Delta t) = y$. It can be seen in figure 4a

177  that the curve of theory is well approximated by the data in different conditional

178  average cases. We can found that, as a consequence of the long-range interaction, the

179  relationship of conditional average tends monotonic, the mean velocity to the other

180  system depends strongly on the velocity in the first system.

181  By the same trick, the analytical expression of the conditional variance of

182  normalized velocity between two-joint systems can be expressed as

$$\sigma^2_{u_2|u_1}(y) = \int_{W_z|y} z^2 p_{u_2|u_1}(z|y) dz. \qquad (15)$$

183  where $\sigma^2_{u_2|u_1}(y)$ is the conditional variance, while it depends on $u_1(t, \Delta t) = y$. It also

184  shows clearly in figure 4b that the curve of theory is well approximated by the data in

185  different conditional variance cases. The figure clearly displays the effect of the

186  long-range interaction: Small and large conditional variance in the other system is

187  more likely to be followed by small and large velocity in the first system, respectively.

188  It must be mentioned that the theory of conditional average and conditional variance

189  presented here is based on the CPDF obtained before. After that, it is possible to

190  calculate other conditional statistical properties base on the obtained CPDF.

191  In addition, we can also analyze the additivity of entropy of two-joint systems

192  based on the obtained CPDF. We can first numerical calculate the entropy of each

193  systems by Shannon method [14] as

$$S_u = -k \int_W p_u(y) \ln[p_u(y)] dy. \qquad (16)$$

194  where k is the Boltzmann parameter, for the sake of convenience, we set k equals 1 in our work. From the two typical currency exchange databases, we can obtain that the entropy of EUR/USD and GBP/USD equal 1.335 and 1.328, respectively. Furthermore, we can obtain the JPDF $p_{u_1,u_2}(y,z)$ of two-joint systems from the obtain CPDF $p_{u_2|u_1}(z|y)$. After that, we can also numerical calculate the entropy of two-joint systems from the obtain JPDF $p_{u_1,u_2}(y,z)$ as

$$S_{u_1,u_2} = -k \iint_W p_{u_1,u_2}(y,z) \ln[p_{u_1,u_2}(y,z)] dy dz. \tag{17}$$

From the two typical currency exchange databases, we can obtain that the joint entropy of EUR/USD and GBP/USD equal 2.616. It is importance noted that the entropy of two-joint systems is not equal to sum of entropy of each system. As is well known, the EUR and GBP are both the European currency, some information of EUR and GBP may overlap each other. As a result, the entropy of two-joint systems is less than the sum of entropy of each system. With the same method, it is possible to calculate other joint statistical physical properties base on the obtained CPDF. It is important to note that the results presented here do not need to know the form of interaction of two-joint systems. Furthermore, this method may be helpful for lead to more accurate calculation of conditional statistical physical properties in joint long-range interaction complex systems, and applicable to the analysis of a wide range of phenomena, including natural, artificial, financial and social complex systems.

To sum up, we have proposed a new method to obtain the conditional statistical physical properties in two-joint long-range interaction complex systems. By analyzing two typical currency exchange database of EUR/USD and GBP/USD as joint systems, we can describe the CPDF $p_{u_2|u_1}(z|y)$ only with six parameters $q_\psi$, $\sigma_\psi$, $r_{\sigma 0}$, $r_{\sigma 1}$, $r_{q0}$ and $r_{q1}$ in all different cases. It is shown clearly that the JPDF between two-joint typical currency exchange systems do not suite the independent principle of multiplying of each system. Moreover, we numerically calculate the conditional average and the conditional variance by using the obtained CPDF $p_{u_2|u_1}(z|y)$. It is important to note that the theory of conditional average and conditional variance presented here is based on the CPDF obtained before. It is possible to calculate other conditional statistical physical properties base on the obtained CPDF. In addition, we also found that, as a consequence of the long-range interaction, the entropy of two-joint systems is less than the sum of entropy of each system. We must also important to note that the results presented here do not need to know the form of interaction of two-joint systems. It may be expected that the further research in this direction will open new perspectives and shed new light on the research of the joint complex systems which having long-range interaction.

**Acknowledgments**

Project supported by the National Natural Science Foundation (No. 11247265, 11005041), the Fujian Provincial Natural Science Foundation (No. 2011J01012), the program for prominent young Talents in Fujian Province University (JA12001), and the Science Research Fund of Huaqiao University (No.11BS207), People's Republic of China.

Figure captions:

Fig.1. The PDF of variables and the CPDF between $\theta(t,\Delta t)$ and $\psi(t,\Delta t)$.

Fig.2. The conditional variance between $\theta(t,\Delta t)$ and $\psi(t,\Delta t)$ for the parameters $r_{\sigma 0}$ and $r_{\sigma 1}$ and the conditional q between $\theta(t,\Delta t)$ and $\psi(t,\Delta t)$ for the parameters $r_{q0}$ and $r_{q1}$.

Fig.3. The CPDF of the velocity between GBP/USD and EUR/USD. Dots and curves correspond to the cases of data and theory, respectively.

Fig.4. The conditional average of the velocity between two typical currency exchange systems and the conditional variance of the velocity between two typical currency exchange systems. Square dots and solid curves correspond to the cases of data and theory, respectively.

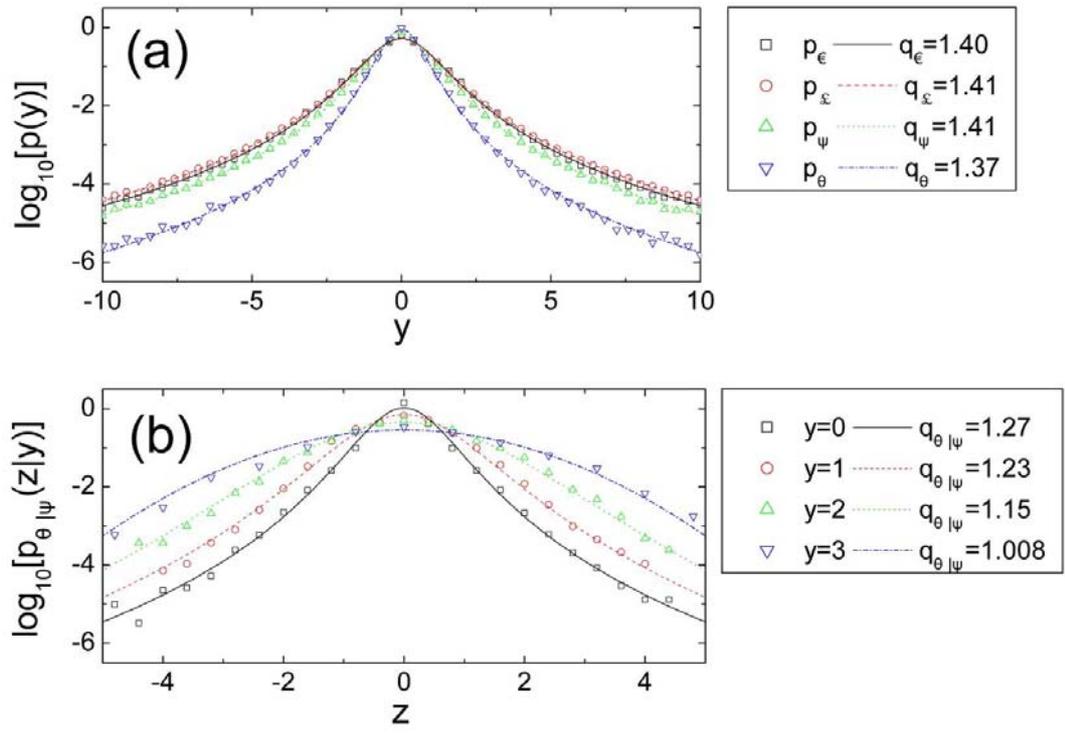

Fig.1.

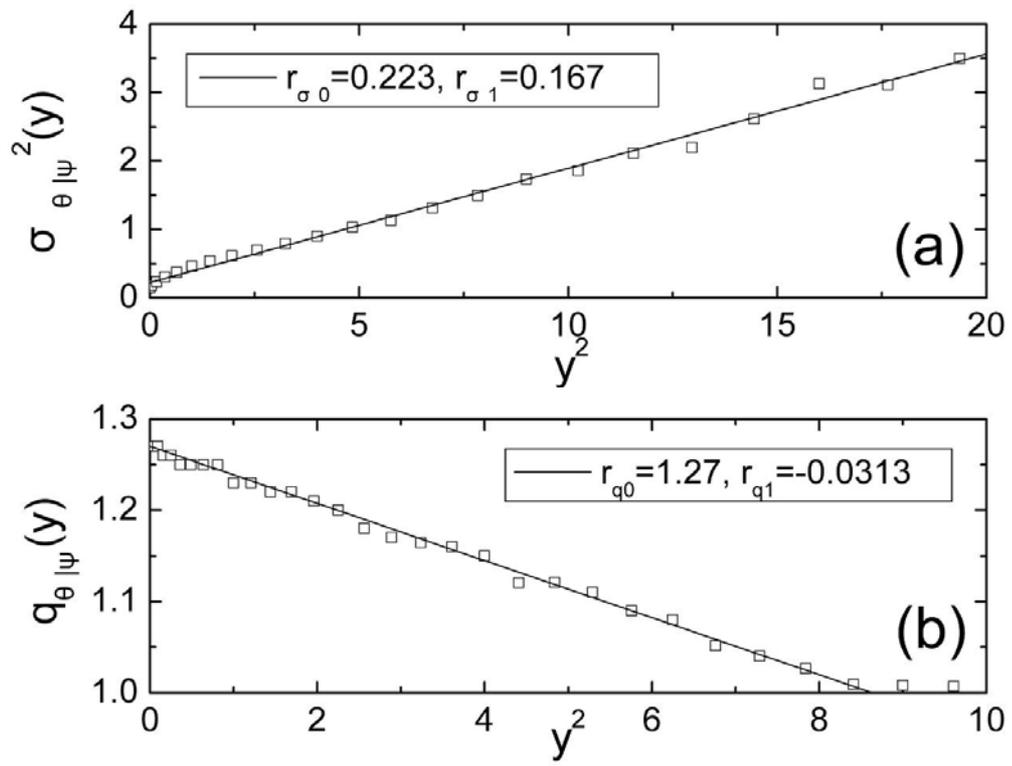

Fig.2.

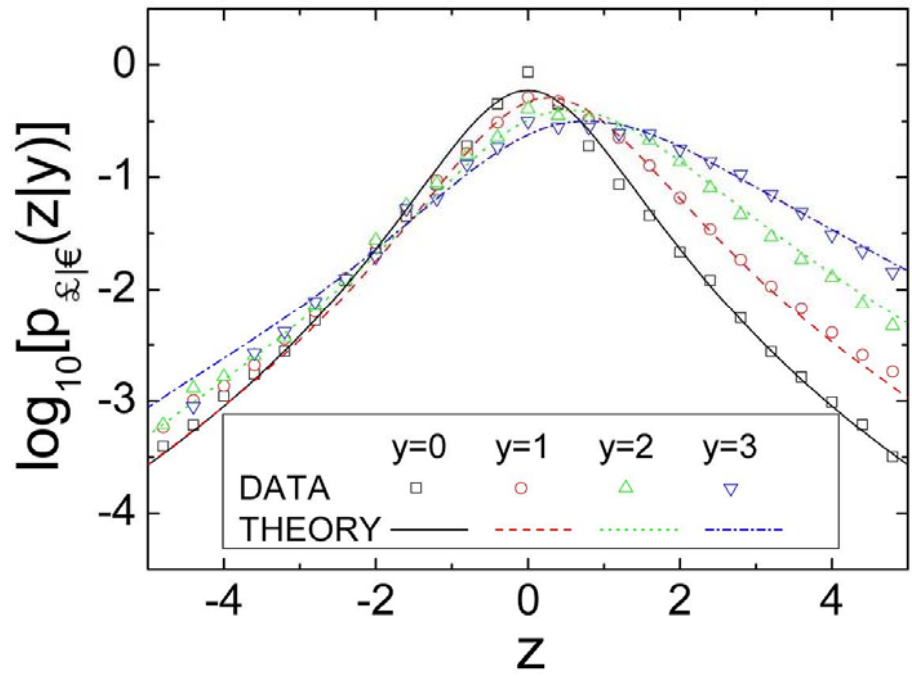

Fig.3.

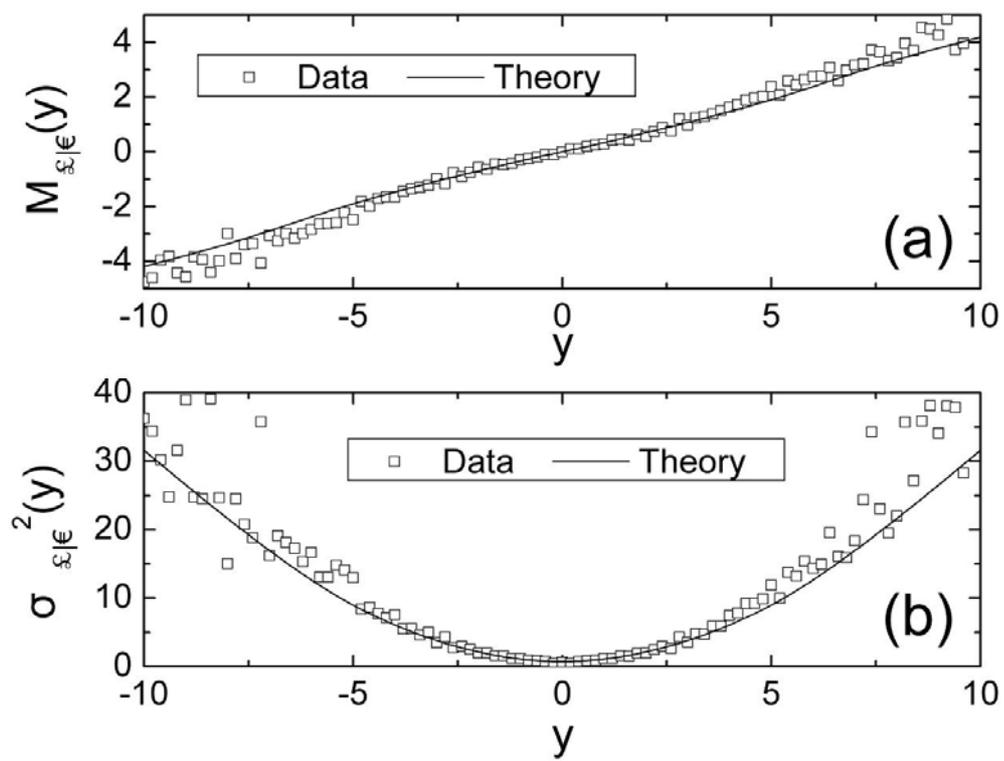

Fig.4.